\documentclass[prb,preprint,superscriptaddress,12pt]{revtex4-1} 

\usepackage{amsmath}  
\usepackage{amsfonts} 
\usepackage{graphicx} 
\usepackage{esvect}
\usepackage{mathtools}
\usepackage{subcaption}
\usepackage{xcolor}
\usepackage{appendix}

\usepackage{siunitx} 
\usepackage{placeins}

\usepackage{float} 

\usepackage{pbox}

\usepackage{mathtools}
\makeatletter
\newcommand{\vast}{\bBigg@{3}}
\newcommand{\Vast}{\bBigg@{3.5}}
\makeatother

\usepackage[mathscr]{euscript}
\DeclareSymbolFont{rsfs}{U}{rsfs}{m}{n}
\DeclareSymbolFontAlphabet{\mathscrsfs}{rsfs}
 
\DeclareMathOperator{\hyperu}{U} 
\DeclareMathOperator{\laguerrel}{L}

\begin{document}


\title{Stress-Induced Mutagenesis \\ Can Further Boost Population Success in Static Ecology}

\author{Kien T. Pham}  
\thanks{These authors contributed equally to this work.}
\affiliation{Department of Aerospace Engineering, School of Transportation Engineering, Hanoi University of Science and Technology, 1 Dai Co Viet Str., Hai Ba Trung District, Hanoi 100000, Vietnam.}

\author{Duc M. Nguyen}
\thanks{These authors contributed equally to this work.}
\affiliation{The College, University of Chicago, Chicago, IL 60637, USA.}

\author{Duy V. Tran}
\affiliation{University of Technology - VNUHCM, 228 Ly Thuong Kiet, Ho Chi Minh 700000, Vietnam.}

\author{Vi D. Ao} 
\affiliation{University of Science - VNUHCM, 227-Nguyen Van Cu Street, 5th District, Ho Chi Minh City, Vietnam.}

\author{Huy D. Tran}
\affiliation{Department of Physics, The Hong Kong University of Science and Technology, Clear Water Bay, Kowloon, Hong Kong, P.R. China.}

\author{Tuan K. Do}
\affiliation{Department of Mathematics, UCLA, Los Angeles, CA 90095-1555, USA.}

\author{Trung V. Phan}
\email{trung.phan@yale.edu}
\affiliation{Department of Molecular, Cellular, and Developmental Biology, Yale University, New Haven, CT 06520, USA}

\begin{abstract}
We have developed a mathematical model that captures stress-induced mutagenesis, a fundamental aspect of pathogenic and neoplastic evolutionary dynamics, on the fitness landscape with multiple relevant genetic traits as a $\mathscrsfs{D}$-dimensional Euclidean space. In this framework, stress-induced mutagenesis manifests as a heterogeneous diffusion process. We show how increasing mutations, and thus reducing exploitation, in a static ecology with fixed carrying capacity and maximum growth rates, can paradoxically boost population size. Remarkably, this unexpected biophysical phenomenon applies universally to any number of traits $\mathscrsfs{D} = 1,2,3,...$
\end{abstract}

\date{\today}

\maketitle

\section{Introduction}

The concepts of exploration and exploitation are often thought of as opposing strategies, with finding the right balance between the two being the key to achieving success, both in the natural world and in human endeavors \cite{gupta2006interplay,uotila2009exploration,berger2014exploration,guisado2017analyzing,alba2005exploration}. Exploitation focuses on maximizing the utilization of existing resources and knowledge to achieve optimal outcomes. In contrast, exploration is the process of searching for better options, which becomes crucial when facing ever-changing challenges. On the other hand, if the challenge remains static, and the optimal solution is well-defined globally, then exploration may be perceived as futile. In biology, genetic mutation functions as the exploration mechanism in the extensive range of all possible combinations of genetic traits (the biological landscape) \cite{wright1932roles,kimura1962probability,kimura1964diffusion}, and a species' success can be gauged by how large its population size is \cite{phan2021it}. For an unchanging environment, it is suggested that for maximum success, the population distribution in the landscape should be as localized as possible around the genetic traits with the highest growth rate, rendering a lower mutation rate more desirable.

However, this claim, as we find out, is fundamentally flawed.

In natural complex and dynamic ecology, where living systems strongly interact with environments \cite{phan2020bacterial}, there exists a coupling between the abstract space of genetic trait variations and outside physical space. We reveal that this coupling can have a significant impact through stress-induced mutagenesis, a common evolutionary mechanism that has been observed not only at the microbial level \cite{bjedov2003stress} but also in cancer cells \cite{fitzgerald2017stress}. This curious mechanism is triggered by high biological stress such as starvation or exposure to antibiotics \cite{imasheva1999environmental,hoffman2000environmental}. For {\it E.coli} bacteria, this process involves the engagement of the SOS response, leading to the induction of low-fidelity error-prone replication polymerases and a sharp increased mutation rate during DNA replication from the normally low value of $D_l \propto 10^{-9}$ to a high rate of $D_h \propto 10^{-5}$ mutations per base pair per generation \cite{cirz2005inhibition, bos2015emergence}. In general, there are many mechanisms by which genetic change can be produced when organisms are under stress \cite{foster2007stress}. The importance of stress-induced mutagenesis, as an essential evolutionary dynamic in the realm of microbiology, has been demonstrated by laboratory studies showing that at least $80\%$ of natural isolates of {\it E.coli} from diverse environments worldwide can exhibit this behavior \cite{pribis2022stress}. For humanity, it poses a significant threat e.g. the direct contribution to the rapid development of drug resistance \cite{zhang2011acceleration,wu2014game,li2021acceleration}, leading to complications in treating infectious diseases. According to the World Health Organization, an estimated 1.27 million deaths worldwide resulted from antibiotic-resistant infections in 2019 \cite{murray2022global}.

Previously, we have demonstrated that more exploration non-stress-induced can lead to better population success for living system in dynamic surrounding \cite{phan2021it}. It also has been reported in the literature that stress-induced mutagenesis can drastically enhance adaptation to ever-changing environment \cite{foster2007stress,ram2014stress}. In this work, we find that the benefit of stress-induced mutagenesis is even more fundamental: the population success $S=B/K$ \cite{phan2021it} can already be boosted in a static ecology where the carrying capacity $K$ \cite{tsoularis2002analysis} and the maximum growth rates $R(\vec{x})$ \cite{gavrilets2018fitness,wright1932roles} for all combinations $\vec{x}$ of $\mathscrsfs{D}$ relevant genetic traits (e.g. scalar-strategies \cite{vincent2005evolutionary}) on the Euclidean-topology landscape ($\vec{x} \in \mathbb{R}^{\mathscrsfs{D}}$) are constants with time. $B = \int d^{\mathscrsfs{D}} \vec{x} b(\vec{x})$, the integration of population density distribution $b(\vec{x})$ over all this abstract space, is the total population in the ecology. Here we present a simple mathematical model in which stress-induced mutagenesis is described on the landscape by a diffusion process \cite{wright1932roles,kimura1962probability,kimura1964diffusion} with growth-dependent diffusivity $D[G]$, where the growth is the same as in the simple logistic-growth model $G=R(1-S)$ \cite{tsoularis2002analysis} and upping mutation rate happens sharply at no-growth $G=0$ (e.g. for {\it E.coli}, this is when cell stops dividing and gets elongated \cite{bos2015emergence,phan2018helical}):
\begin{equation}
D[G>0] = D_l \ , \ D[G<0] = D_h \geq D_l \ \ ,
\label{stress_induced}
\end{equation}
analogous to the realistic response \cite{cirz2005inhibition, bos2015emergence}. Let us define:
\begin{equation}
\epsilon=(D_h-D_l)/D_l \geq 0 \ , 
\label{define_epsilon}
\end{equation}
then $\epsilon$ can be used as a metric to quantify the strength of stress-induced mutagenesis, which increases when this effect becomes stronger. We show, with analytical and numerical investigations, that the success $S$ at stationary-state goes up monotonically with the strength $\epsilon$. Remarkably, we also can prove the universality of this finding for any number of traits $\mathscrsfs{D} \in \mathbb{N}$, indicating the generality of our claim. Our results highlight an underdeveloped sector of theoretical evolutionary dynamics, which is common in the world of cells and can exhibit such strange and counter-intuitive phenomena, and emphasize the potential of ultilizing this mathematical framework to uncover the emergent complexities of evolution.

\section{Theoretic Considerations for Contributions from Stress-Induced Mutagenesis}

In Section \ref{toy-model}, we introduce our mathematical toy-model for evolutionary dynamics, which incorporates genetic diversity with varying fitness, growth constraints that reflect ecological limitations on resources and physical competition, and, most importantly, stress-induced mutagenesis. Moving to Section \ref{without}, we demonstrate that without stress-induced mutagenesis $\epsilon=0$, increasing the exploration rate by uniformly raising the mutation rate $D_l=D_h \uparrow$ leads to a decrease in population success $S \downarrow$. Finally, in Section \ref{with}, we examine the effect of stress-induced mutagenesis $\epsilon>0$, which promotes exploration only for unfit combinations of genetic traits with a high mutation rate $D_h \uparrow$. We reveal how this evolutionary mechanism can substantially and universally boost population success $S \uparrow$, even rescuing the population from total extinction.

\subsection{An Integro-differential equation for Stress-Induced Mutagenesis \label{toy-model}} 

Mutation can be regarded as a stochastic process involving random-walk on the abstract space of genetic variations \cite{kimura1964diffusion,phan2021it}. In this space, each combination of traits is represented by a point $\vec{x}$ ($\vec{x} \in \mathbb{R}^\mathscrsfs{D}$ for $\mathscrsfs{D}$ evolutionary scalar strategies \cite{vincent2005evolutionary}) and exhibits unchanging maximum growth rate $R(\vec{x})$ in static ecology. The distribution density $b(\vec{x},t)$ of the population in this landscape can be modeled by the following Fokker-Planck equation \cite{risken1996fokker}:
\begin{equation}
\partial_t b = \nabla^2 \left( D b \right) + R b \ ,
\label{fokker_planck}
\end{equation}
where the diffusivity $D$ represents the speed of mutations.

In every ecological system, unlimited population growth is unsustainable due to limited resources in the environment. The logistic model of population growth is a fundamental description that accounts for this limitation \cite{tsoularis2002analysis}. In such models, competition for resources results in a carrying capacity $K$ \cite{getz1991unified,getz1994metaphysiological}, which can be incorporated in the mathematical framework by modifying Eq. \eqref{fokker_planck} into an integro-differential equation:
\begin{equation}
\partial_t b = \nabla^2 \left( D b \right) + G b \ , 
\label{fokker_planck_logistic}
\end{equation}
where $G = R\left( 1-S \right)$ is the growth rate, $S=B/K$ is the population success, and $B$ is the population size which can be calculated as the integration of population density distribution $B(t) = \int d^{\mathscrsfs{D}} \vec{x} b(\vec{x},t)$. To include a realistic representation of stress-induced mutagenesis, we use the two-state growth-dependent diffusivity $D[G]$ as argued in \eqref{stress_induced}. To increase the veracity of the model, it is necessary to apply a generalized Richards law \cite{richards1959flexible} to regulate growth, where the effective exponent varies with population size \cite{swartz2022seascape,tran2022extinction}. 

\begin{figure}[ht]
\centering
\includegraphics[width=\textwidth,keepaspectratio]{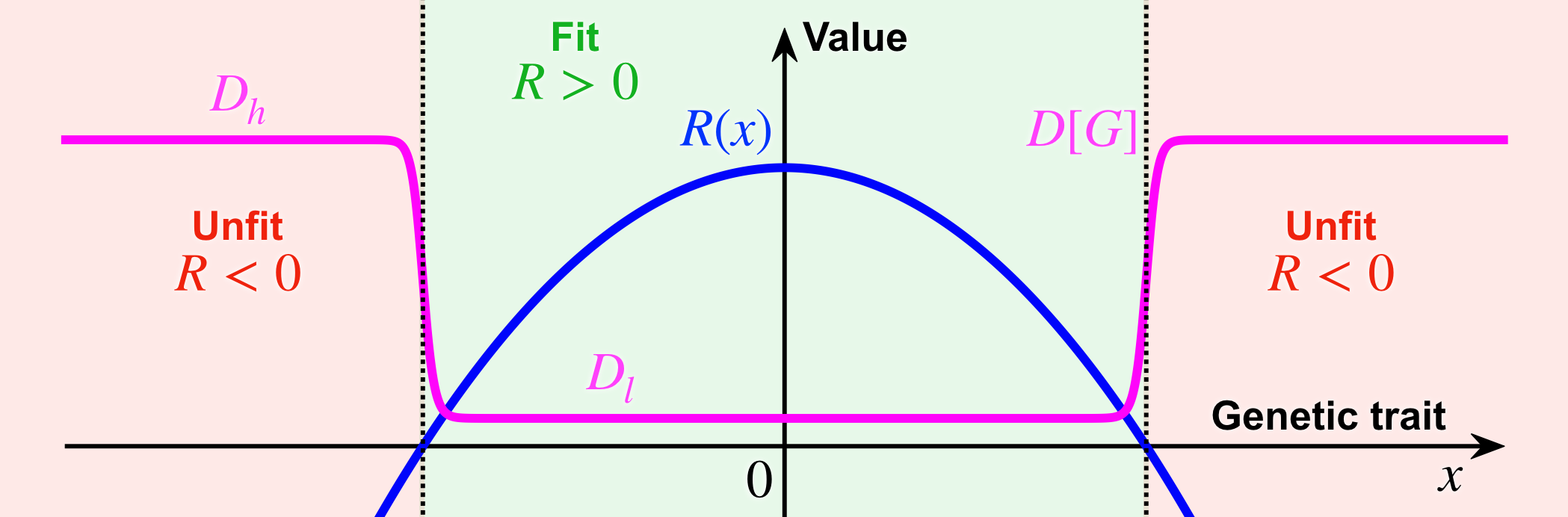}
\caption{The landscape of our toy-model, illustrated for a single genetic trait $\mathscrsfs{D}=1$. The effective diffusivity $D[G]$ shifts from a low value $D_l$ to a high value $D_h$ when the growth rate $G \propto R$ switches sign.}
\label{fig1}
\end{figure}

Here comes the ``toy'' parts of our toy-model. For the sake of simplicity we assume that the fitness landscape is symmetric around the position of the optimal combinations of traits (taken to be the origin) $\vec{x}_{op}=0$:
\begin{equation}
R(\vec{x}) \approx R_0 \left[ 1 - \left(\frac{\vec{x}}{\lambda}\right)^2 \right] \ .
\label{toy_growth}
\end{equation}
The further away from the optimal combinations, the lower the fitness. Negative growth means the number of deaths exceeds the number of births, a crucial indicator on the fitness landscape showing the impact of natural selection  \cite{mccandlish2022visualizing}. Since $G \propto R$, the characteristic sharp transition from $D_l$ to $D_h$ in $D[G]$ at no-growth also happens right at $R=0$, where $|\vec{x}|=\lambda$. We call $|\vec{x}|<\lambda$ the {\it fit region} where $R>0$, and $|\vec{x}|>\lambda$ the {\it unfit region} where $R<0$ \cite{phan2021it}. The basics of our model are conveyed in Fig. \ref{fig1}.

We can change the spatial coordinates from Cartesian $\vec{x}$ to generalized-spherical coordinates $\left( r, \Omega_{\mathscrsfs{D}-1}\right)$ then subsequently apply a rescaling $
\rho = r/\lambda$, so that the dimensionless radial measure $\rho<1$ stands for the fit region and $\rho>1$ corresponds to the unfit region. We can also rescale the population measure $\beta=b\lambda^{\mathscrsfs{D}}/K$, so that $S=\int \rho^{\mathscrsfs{D}-1}d\rho d\Omega_{\mathscrsfs{D}-1} \beta$. Let us consider a radial symmetric distribution $\beta = \beta(\rho,t)$, Eq. \eqref{fokker_planck_logistic} becomes:
\begin{equation}
\partial_t \beta = \frac1{\rho^{\mathscrsfs{D}-1}} \partial_\rho \Bigg[ \rho^{\mathscrsfs{D}-1} \partial_\rho  \left( 
 \Big[ 1 + \epsilon\Theta(\rho-1) \Big] \beta \right) \Bigg] + \Xi \left( 1-S \right) \left( 1-\rho^2 \right) \beta \ ,
\label{fokker_planck_logistic_dimless}
\end{equation}
where $\Theta(\zeta)$ is the Heaviside function in which $\Theta(\zeta<0)=0$ and $\Theta(\zeta>0)=1$ \cite{heaviside1893electromagnetic}, and $\Xi = R_0 \lambda^2/D_l$. The larger fit region size $\lambda \uparrow$, the smaller the parameter $\Xi \downarrow$. Same goes for higher maximum optimal growth rate $R_0 \uparrow$ and lower diffusivity $D_l \downarrow$.

Through the solution $\beta = \beta_{st}(\rho)$ for the stationary state of Eq. \eqref{fokker_planck_logistic_dimless} by setting $\partial_t \beta=0$ and the subsequent determination of the resulting $S_{st}$, we can gain new insights into the effect of stress-induced mutagenesis at population level.

\subsection{Without Stress-Induced Mutagenesis \label{without}}

When there is no stress-induced mutagenesis, $\epsilon=0$ and Eq. \eqref{fokker_planck_logistic_dimless} simplifies:
\begin{equation}
0 = \frac1{\rho^{\mathscrsfs{D}-1}} \partial_\rho \left( \rho^{\mathscrsfs{D}-1} \partial_\rho  \beta_{st}  \right) + \omega_l^2 \left( 1-\rho^2 \right) \beta_{st} ,
\label{no_stress_induced}
\end{equation}
in which we define:
\begin{equation}
\omega_l^2 = \Xi (1-S_{st}) = \frac{R_0\lambda^2}{D_l}(1-S_{st}) \ .
\label{omega_l}
\end{equation}
The general solution for Eq. \eqref{no_stress_induced} are given by the linear superposition of a Tricomi confluent hypergeometric-function $\hyperu(...)$ and a Laguerre polynomial $\laguerrel(...)$\cite{abramowitz1948methods}:
\begin{equation}
\beta_{st} = e^{-\frac12 \omega_l \rho^2} \left[ \beta_1 \  \hyperu\left( \frac{\mathscrsfs{D}-\omega_l}4 , \frac{\mathscrsfs{D}}2,\omega_l \rho^2\right) \ + \ \beta_2 \ \laguerrel \left(-\frac{\mathscrsfs{D}-\omega_l}4, \frac{\mathscrsfs{D}}2-1,\omega_l \rho^2 \right) \right] \ .
\label{constant_diff_general_sol}
\end{equation}
The conditions that a physical $\beta_{st}$ has to satisfy are: (i) has no kink at $\rho=0$; (ii) asymptotically goes to $0$ at far-away infinity $\rho \rightarrow 0$; and (iii) can never be negative anywhere. These requirements uniquely determine the shape of $\beta_{st}$ and the value of $\omega_l$:
\begin{equation}
\beta_{st} \propto e^{-\frac12 \omega_l \rho^2} \ , \ \omega_l = \mathscrsfs{D} \ .
\label{constant_diff_sol}
\end{equation}
Therefore, the stationary population success is given by:
\begin{equation}
S_{st} = 1 - \frac{\omega_l^2}{\Xi} = 1 - \frac{D_l \mathscrsfs{D}^2}{R_0 \lambda^2}\ ,
\label{no_stress_induced_success}
\end{equation}
which decreases as the diffusivity $D_l \uparrow$ increases. This finding corresponds to the expectation that more exploration results in lower success for static ecology. We note that it is also the same consequence with when there are more traits $\mathscrsfs{D} \uparrow$.

We show how this intuitive claim fails in the next Section, once effect of stress-induced mutagenesis $\epsilon > 0$ is considered.

\subsection{With Stress-Induced Mutagenesis \label{with}}

\begin{figure}[ht]
\centering
\includegraphics[width=\textwidth,keepaspectratio]{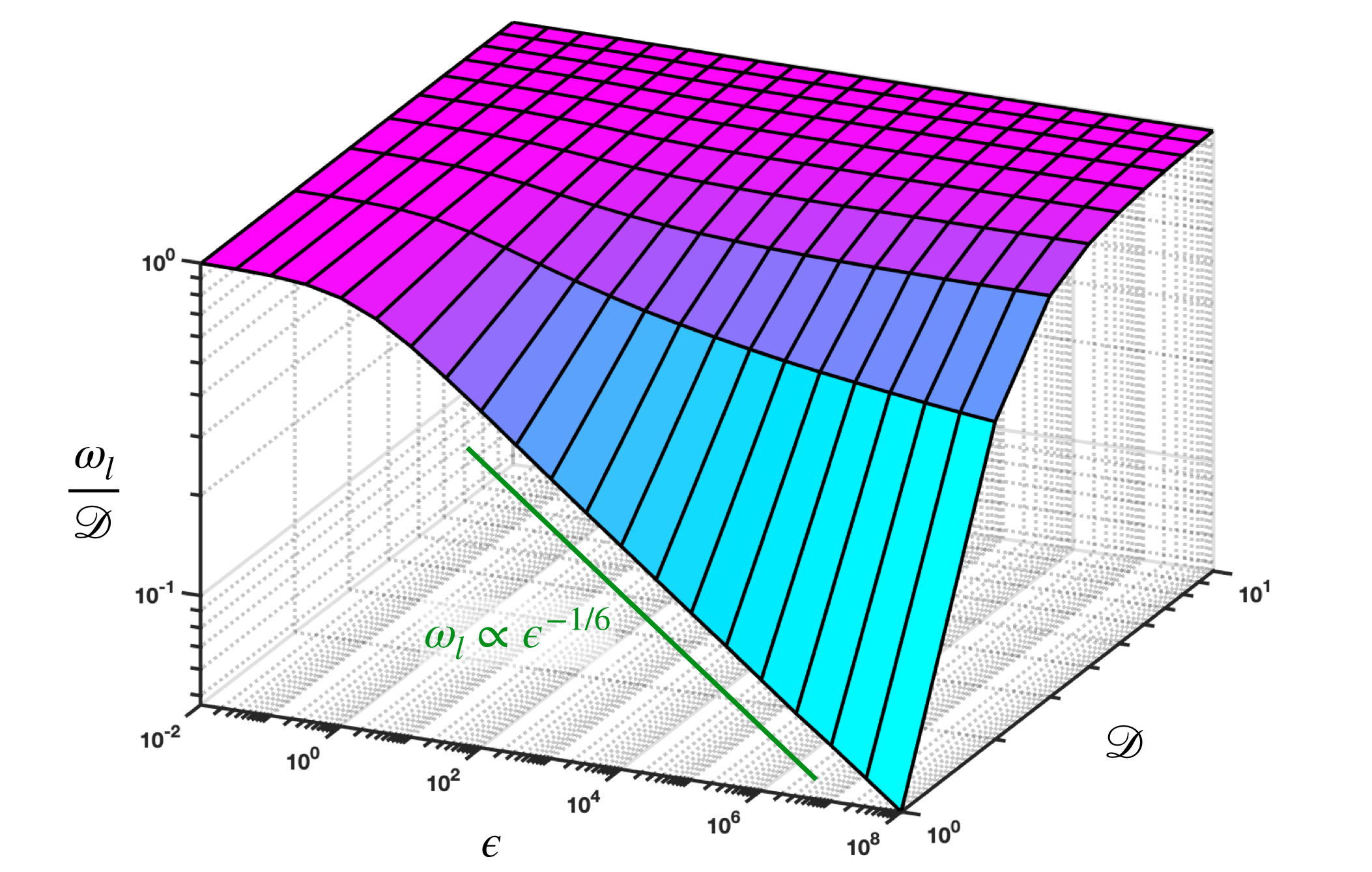}
\caption{The surface plot $\omega_l(\mathscrsfs{D},\epsilon)$, in which the value of $\omega_l$ always decrease as $\epsilon$ increases for every natural dimensionality $\mathscrsfs{D}$. The parameter space of investigation spans from $\epsilon=10^{-2}$, $\mathscrsfs{D}=10^{0}$ to $\epsilon=10^{8}$, $\mathscrsfs{D}=10^{1}$ (see Appendix \ref{detail_intersection}).}
\label{fig2}
\end{figure}

To deal with Eq. \eqref{fokker_planck_logistic_dimless} when stress-induced mutagenesis is in effected $\epsilon>0$, we have to stitch the two solutions $\beta_{st}^{(l)}$ and $\beta_{st}^{(h)}$ of the form given in Eq. \eqref{constant_diff_general_sol} together, in which $\beta_{st}^{(l)}$ is valid for the fit region $\rho<1$ and $\beta_{st}^{(h)}$ is applicable for the unfit region $\rho>1$. Because in general $\hyperu(...)$ has a kink at $\rho=0$ and $\laguerrel(...)$ diverges when $\rho\rightarrow \infty$, we are looking for a physical $\beta_{st}$ of the following ansatz: 
\begin{equation}
\begin{split} 
\beta_{st} \Big|_{\rho<1} & = \beta^{(l)}_{st} \propto e^{-\frac12 \omega_l \rho^2} \ \laguerrel \left(-\frac{\mathscrsfs{D}-\omega_l}4, \frac{\mathscrsfs{D}}2-1,\omega_l \rho^2 \right) \ ,
\\
\beta_{st} \Big|_{\rho>1} &= \beta^{(h)}_{st} \propto e^{-\frac12 \omega_h \rho^2} \  \hyperu\left( \frac{\mathscrsfs{D}-\omega_h}4 , \frac{\mathscrsfs{D}}2,\omega_h \rho^2\right) \ .
\label{constant_diff_general_sol_2sides}
\end{split}
\end{equation}
where we define $\omega_h$ as a function of $\omega_l$, using Eq. \eqref{define_epsilon}:
\begin{equation}
\omega_h^2 = \left(\frac{\omega_l}{\sqrt{1+\epsilon}} \right)^2 = \frac{R_0\lambda^2}{D_h}(1-S_{st}) \ .
\label{omega_h}
\end{equation}
For regular flux-continuation $\nabla \left( D\beta \right)$ at the transition interface $\rho=1$, we need to match up Eq. \eqref{constant_diff_general_sol_2sides} such that:
\begin{equation}
D_l \beta_{st} \Big|_{\rho=1^-} = D_h \beta_{st} \Big|_{\rho=1^+} \ , \  D_l \partial_\rho \beta_{st} \Big|_{\rho=1^-} = D_h \partial_\rho \beta_{st} \Big|_{\rho=1^+} \ . 
\label{matching}
\end{equation}
After performing some algebraic manipulations, it follows that the value of $\omega_l$ must satisfy:
\begin{equation}
\omega_l \left[ 1 + \frac{ 2 \ \laguerrel \left(-\frac{\mathscrsfs{D}-\omega_l}4-1,\frac{\mathscrsfs{D}}2,\omega_l\right) }{\laguerrel \left(-\frac{\mathscrsfs{D}-\omega_l}4,\frac{\mathscrsfs{D}}2-1, \omega_l \right) } \right]
 = \omega_h \left[ 1 + \frac{ \left( \mathscrsfs{D}-\omega_h\right) \ \hyperu \left( \frac{\mathscrsfs{D}-\omega_h}4+1,\frac{\mathscrsfs{D}}2+1,\omega_h\right)}{2 \ \hyperu\left( \frac{\mathscrsfs{D}-\omega_h}4 , \frac{\mathscrsfs{D}}2,\omega_h \right)} \right] \ . 
\label{matching_for_omega_l}
\end{equation}
This equation can be solved numerically, as we describe in Appendix \ref{detail_intersection}, giving us the surface function of $\omega_l(\mathscrsfs{D},\epsilon)$ as shown in Fig. \ref{fig2}:
\begin{equation}
\boxed{
\omega_l(\mathscrsfs{D},0) = \mathscrsfs{D} \ > \ \omega_l(\mathscrsfs{D},\epsilon) \Big|_{\epsilon>0} \ \ \forall \mathscrsfs{D} \in \mathbb{N} \ .
}
\label{key_result}
\end{equation}
This is our key result. Since $S_{st}$ and $\omega_l$ are monotonically opposed from the definition in Eq. \eqref{omega_l}, this means it is universal for any number of relevant genetic traits $\mathscrsfs{D} = 1,2,3, ...$ that stress-induced mutagenesis $\epsilon \uparrow$ always boosts up population success $S_{st} \uparrow$. For $\mathscrsfs{D}=1$ we get the asymptotic behavior $\omega_l \propto \epsilon^{-1/6}$ at large value of $\epsilon \gg 1$, which is the regime of stress-induced mutagenesis observed in real living {\it E.coli} bacteria $\epsilon \sim 10^4$ \cite{cirz2005inhibition,bos2015emergence}. In that same regime, for $\mathscrsfs{D}=2$ the value of $\omega_l$ very gradually approaches $0$ and for $\mathscrsfs{D}>2$ the value of $\omega_l$ converges to finite constants. The mathematical details regarding these analytical considerations can be found in Appendix \ref{detail_math}.

\begin{figure}[ht]
\centering
\includegraphics[width=\textwidth,keepaspectratio]{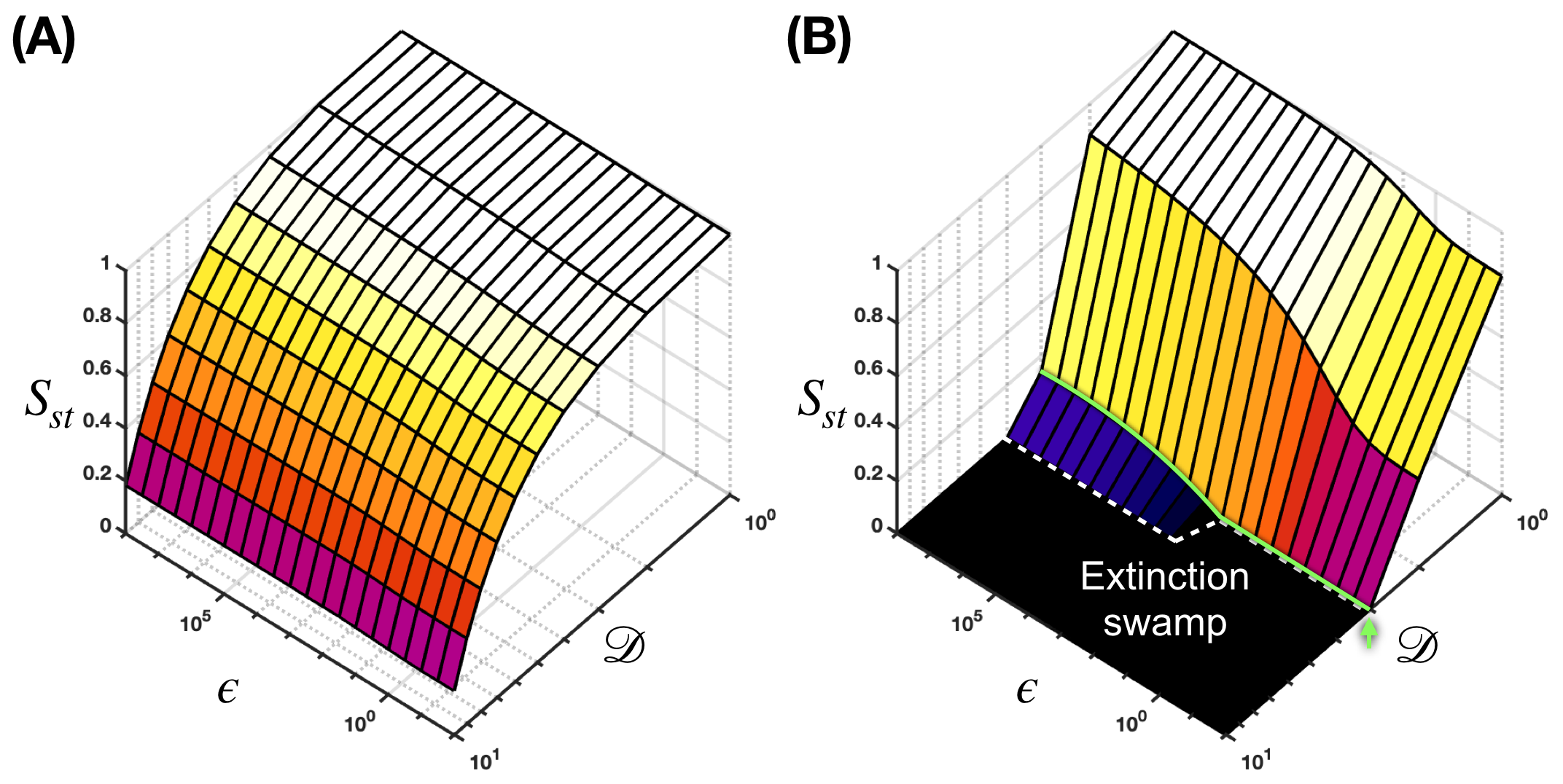}
\caption{The surface plots for $S(\mathscrsfs{D},\epsilon)$, investigated in the parameter space spans from $\epsilon=10^{-2}$, $\mathscrsfs{D}=10^{0}$ to $\epsilon=10^{8}$, $\mathscrsfs{D}=10^{1}$, in which: (A) $\Xi = 120$, the stress-induced mutagenesis effect seems negligible at low number of relevant genetic traits $\mathscrsfs{D}$. (B) $\Xi = 6$, the stress-induced mutagenesis effect is substantial at low number of traits $\mathscrsfs{D}$. We can observe an extinction swamp here. We highlight $S(8,\epsilon)$ by a green curve, which is deep in the swamp until $\epsilon \gtrsim 10^4$.}
\label{fig3}
\end{figure}

Although the impact of stress-induced mutagenesis on $S_{st}$ may appear negligible for biological systems having a wide fit region $\lambda \uparrow$  (associated with a large value of $\Xi \uparrow$, as shown in Fig. \ref{fig3}A for $\Xi=120$), it is of substantial significance when the fit region is narrow $\lambda \downarrow$ (resulting in a minuscule value of $\Xi \downarrow$).  To illustrate this point, consider the case of $\Xi=6$, where Fig. \ref{fig3}B displays an {\it extinction swamp} with no possibility for any non-zero stationary population to thrive \cite{phan2021it,wang2022robots}. For $\mathscrsfs{D}=8$, as $\epsilon$ approaching $\sim 10^4$ (similar to {\it E.coli} \cite{cirz2005inhibition,bos2015emergence}) the population can escape the swamp and become sustainable. In other words, stress-induced mutagenesis has come to the rescue!

\begin{figure}[ht]
\centering
\includegraphics[width=\textwidth,keepaspectratio]{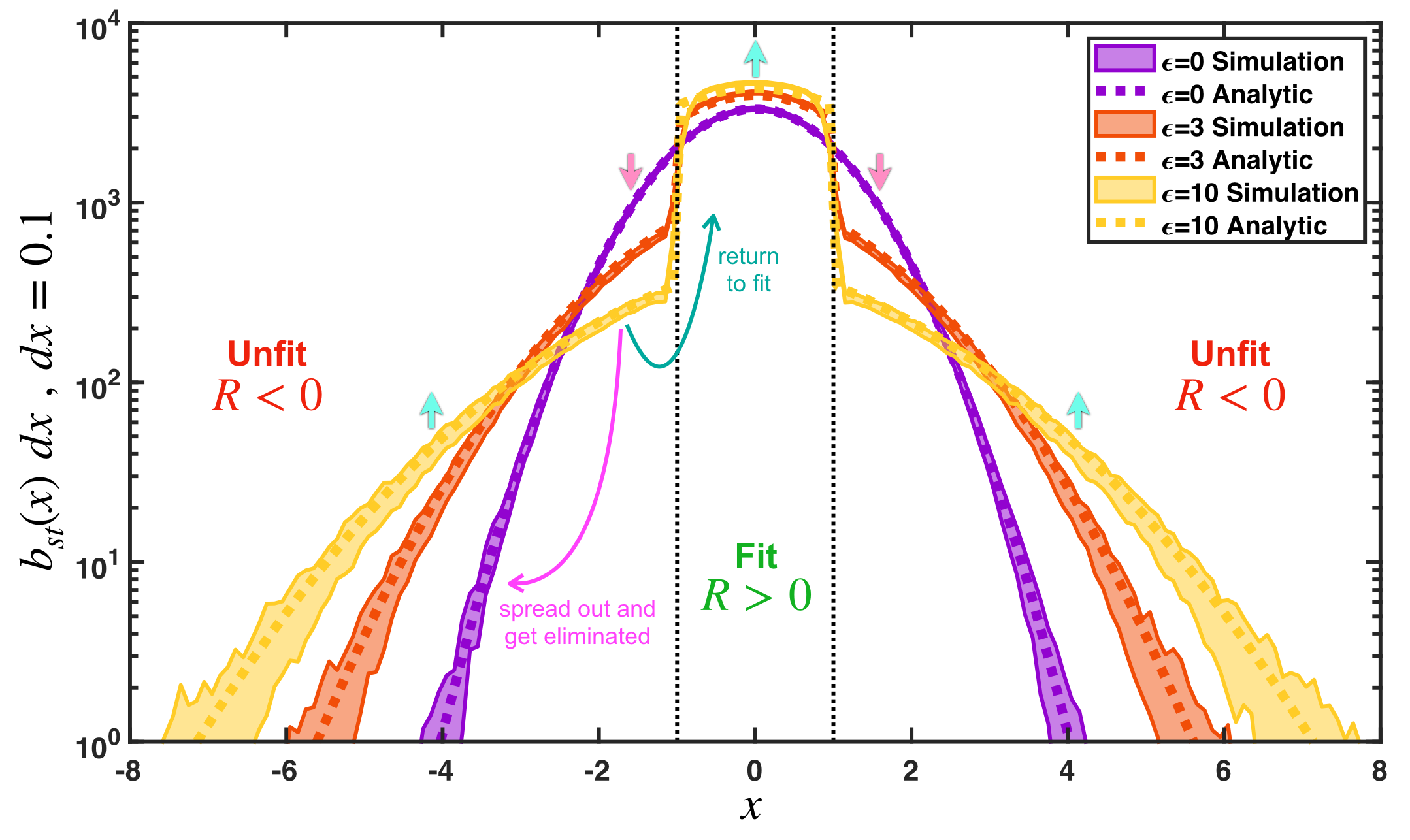}
\caption{We compare our analytic results following from Eq. \eqref{constant_diff_general_sol_2sides} and Eq. \eqref{matching_for_omega_l} with the stationary states obtained from an agent-based simulation (see Appendix \ref{detail_simulation}). Increasing the strength of stress-induced mutagenesis leads to an increase in the number of fit individuals and a wider dispersion of the unfit subpopulation across the landscape.}
\label{fig4}
\end{figure}

In order to gain insights into the emergence of boosted population success in response to stress-induced mutagenesis, it is crucial to understand the distribution of the population across the fitness landscape. To that end, we show in Fig. \ref{fig4} the analytic results obtained from equations \eqref{constant_diff_general_sol_2sides} and \eqref{matching_for_omega_l}, which has also been verified through an agent-based random-walk simulation as described in detail in Appendix \ref{detail_simulation}. Rapid dispersion of the unfit subpopulation across the fitness landscape alleviates the burden of resource limitation on fit individuals, enhancing the effective fitness of the population and increasing the number of fit individuals compared to the absence of stress-induced mutagenesis.

\section{Discussion}

The renowned adage, ``nothing makes sense in biology except in the light of evolution \cite{dobzhansky1973nothing},'' suggests that natural selection favors evolutionary adaptations that assist in optimizing the success of living communities within their ecological niche \cite{morris2017bacterial}. Nevertheless, the existence of stress-induced mutagenesis, which may have been widely present at the dawn of life \cite{NRC1990the,bjedov2003stress}, poses a baffling question: how does this up-exploration mechanism enhance population success in seemingly unchanging environments, which would not be so uncommon due to the rapid doubling time of cells?

To offer an insightful response to this intriguing puzzle, we have formulated a mathematical framework that draws upon analytical components from standard fitness landscape models, incorporating extensions to account for the effects of adaptive mutagenesis and ecological influence. Our aim is geared toward providing a deeper quantitative understanding of stress-induced mutagenesis. With the model we have developed, not only do we reaffirm the conventional wisdom that an increase in exploration via a uniformly elevated mutation rate decreases population success as derived in Eq. \eqref{no_stress_induced_success}, but we also show that focusing on enhancing the mutation rate of an unfit phenotype sub-population -- the essential feature of stress-induced mutagenesis observed in living systems \cite{cirz2005inhibition,bos2015emergence} -- can significantly and universally boost up population success as implied by Eq. \eqref{key_result} and Fig. \ref{fig2}.

Stress-induced mutagenesis drives the evolution of bacterial pathogens, promoting the emergence of dangerous viral variants and antibiotic resistance \cite{taddei1997stress}. This carries significant implications for infectious disease and biology. As an illustrative example in recent times, stress-induced mutagenesis has been discovered to have a notable impact on the evolution of the SARS-CoV-2 virus and the emergence of novel variants \cite{kemp2021sars}. Furthermore, antibiotic misuse has led to the development of antibiotic-resistant bacteria, and this evolutionary mechanism has accelerated the process by promoting mutations that confer resistance \cite{cirz2005inhibition}. Additionally, stress-induced mutagenesis may contribute to cancer treatment resistance, highlighting the need for further studies \cite{fitzgerald2017stress}. Our paper has provided a mathematical framework for comprehending stress-induced mutagenesis, which, as we have demonstrated, can potentially unveil novel and unforeseen emergent population dynamics. Including better representation for ecological influence in the framework here, beside just a mean-field carrying capacity, is also a good territory to explore \cite{wang2021emergent,phan2021bootstrapped}. By laying down a rigorous quantitative foundation, we can probe further into the behavior of adaptive mutagenesis under diverse scenarios, proffering rigorous insights into this curious sector of evolution. We anticipate further investigations of our approach to deepen the understanding of stress-induced mutagenesis, such as why a stochastic dynamical environment can effectively target highly mutating agents and whether that can be utilized for pathogenic and neoplastic treatments \cite{day2022evolving,wang2022robots,levine2022let}.

\appendix
\section{The Estimation of $\omega_l / \mathscrsfs{D}$ as a Function of $\epsilon$ \label{detail_intersection}}

For each natural dimensionality $\mathscrsfs{D}=1,2,3,...$, to find the value $\omega_l$ for every given $\epsilon$, we numerically investigate the LHS and RHS of Eq. \eqref{matching_for_omega_l} as functions of $\omega_l$ and estimate their intersection inside the range $[0,\mathscrsfs{D}]$. 

The code for this estimation is available from the corresponding author upon request. We have executed it on MatLab 2020b.

\section{Some Calculations for Approximations and Asymptotes of $\omega_l \left( \mathcal{D},\epsilon\right)$ \label{detail_math}} 

As mentioned, we are interested in finding the solution of $ \omega_{l} $ for $ \epsilon\gg1 $, since this is the parameter regime observed in the stress-induced mutagenesis of real living bacteria \cite{cirz2005inhibition,bos2015emergence}. By noting that $ \omega_{h}\to0 $ as $ \epsilon\to\infty $, we use the following confluent hypergeometric expansion:
\begin{equation}
    \hyperu \left(a, b, q \right) \xrightarrow{q \rightarrow 0} \frac{\Gamma \left( 1 - b \right)}{\Gamma \left( a - b + 1 \right)} + \frac{\Gamma \left( b - 1\right)}{\Gamma\left( a \right)} q^{1 - b},
\end{equation}
to approximate the RHS of Eq. \eqref{matching_for_omega_l}. This expansion exhibits different behaviors in three different cases: $ \mathscrsfs{D}=1 $, $ \mathscrsfs{D}=2 $, $ \mathscrsfs{D}\geq3 $.

For $ \mathscrsfs{D}=1 $ and $ \mathscrsfs{D}=2 $, we see from Appendix \ref{detail_intersection} that the value of $\omega_l$ gradually approaches 0 as $ \epsilon\to\infty $. For this reason we would like to expand the LHS of Eq. \eqref{matching_for_omega_l} in the limit $ \omega_{l}\to0^{+} $. The leading term in this expansion is:
\begin{equation}
2\alpha(\mathscrsfs{D}) \omega_l^2
\label{RHS_D=1_2}
\end{equation}
where:
\begin{equation}
 \begin{split}
        \alpha (\mathscrsfs{D}) =& \frac{\frac{d}{d\zeta} L \left(\zeta,\frac{\mathscrsfs{D}}2,0 \right)\Big|_{\zeta = -\frac{\mathscrsfs{D}}{4} - 1} - 4\laguerrel \left( -\frac{\mathscrsfs{D}}{4}-2, \frac{\mathscrsfs{D}}{2} + 1, 0 \right)}{4\laguerrel \left( -\frac{\mathscrsfs{D}}{4}, \frac{\mathscrsfs{D}}{2} - 1, 0 \right)}\\
& \ \ \ \ \ \ \ \ -\dfrac{\laguerrel \left( -\frac{\mathscrsfs{D}}{4}-1, \frac{\mathscrsfs{D}}{2} , 0 \right) \left[\frac{d}{d\zeta} L \left(\zeta,\frac{\mathscrsfs{D}}2 - 1,0 \right)\Big|_{\zeta = -\frac{\mathscrsfs{D}}{4}} 
 - 4\laguerrel \left( -\frac{\mathscrsfs{D}}{4}-1, \frac{\mathscrsfs{D}}{2} , 0 \right) \right]}{4\laguerrel^2 \left( -\frac{\mathscrsfs{D}}{4}, \frac{\mathscrsfs{D}}{2} - 1, 0 \right)} \ .
\end{split}
\end{equation}

For large $\mathscrsfs{D}$ ($\mathscrsfs{D} \geq 3$), the value of $\omega_l$ converges to finite constants near $\mathscrsfs{D}^-$, which means we can take asymptotic expansion for $(\omega_l - \mathscrsfs{D})$. For that reason, the LHS of Eq. \eqref{matching_for_omega_l} becomes:
\begin{equation}
    \omega_l \left[ 1 + \frac{1}{2} (\omega_l - \mathscrsfs{D}) \frac{d}{d\zeta} \laguerrel \left(\zeta,\frac{\mathscrsfs{D}}2 ,\mathscrsfs{D} \right)\Big|_{\zeta = -1} \right].
\label{RHS_D=3}
\end{equation}

\subsection{Estimation of $\omega_l (\epsilon)$ for $\mathscrsfs{D} = 1$}

For $\mathscrsfs{D} = 1$, following Eq. \eqref{kummerasymptote} we have:
\begin{equation}
    \frac{ \hyperu \left( \frac{\mathscrsfs{D}-\omega_h}4+1,\frac{\mathscrsfs{D}}2+1,\omega_h\right)}{ 
\hyperu\left( \frac{\mathscrsfs{D}-\omega_h}4 , \frac{\mathscrsfs{D}}2,\omega_h \right)} \Bigg|_{\mathscrsfs{D} = 1} \xrightarrow{\omega_h \to 0}  \frac{\Gamma \left(3/4 \right)}{\Gamma \left(5/4 \right)} \omega_h^{-1/2} \ .
\label{kummerasymptote}
\end{equation}
Together with Eq. \eqref{RHS_D=1_2}, Eq. \eqref{matching_for_omega_l} thus becomes:
\begin{equation}  
2\alpha(1) \omega_l^2 \approx \frac{\mathscrsfs{D}}{2} \left[ \frac{\Gamma \left(3/4 \right)}{\Gamma \left(5/4 \right)} \omega_h^{-1/2}\right] \omega_h 
\ \ \Longrightarrow \ \ \omega_l (\epsilon) = \left[ \frac{1}{4\alpha(1)} \frac{\Gamma \left(3/4 \right)}{\Gamma \left(5/4 \right)}\right]^{2/3} \epsilon^{-1/6} \propto \epsilon^{-1/6} \ .
\end{equation}

\subsection{Estimation of $\omega_l (\epsilon)$ for $\mathscrsfs{D} = 2$}

For $\mathscrsfs{D} = 2$, we use Eq. \eqref{kummerasymptote} to obtain:
\begin{equation}
    \frac{ \hyperu \left( \frac{\mathscrsfs{D}-\omega_h}4+1,\frac{\mathscrsfs{D}}2+1,\omega_h\right)}{ 
\hyperu\left( \frac{\mathscrsfs{D}-\omega_h}4 , \frac{\mathscrsfs{D}}2,\omega_h \right)}\Bigg|_{\mathscrsfs{D} = 2} \xrightarrow{\omega_h \to 0} -\frac{2}{\omega_h \ln \omega_h} \ .
\end{equation}
Plugging this in the LHS and Eq. \eqref{RHS_D=1_2} in the RHS of Eq. \eqref{matching_for_omega_l}, we can arrive at:
\begin{equation}
    2 \alpha(2) \omega_l^2 = \omega_h \left(1 - \frac{\mathscrsfs{D} }{\omega_h \ln \omega_h} \right) \ \ \Longrightarrow \ \ \frac{\omega_l^2}{\epsilon} \ln \frac{\omega_l^2}{\epsilon} = -\frac{2}{\alpha(2) \epsilon}
\end{equation}
The solution of this equation can be represented by the Lambert-W function, which application has been found in protein physics \cite{khuri2021protein}:
\begin{equation}
\omega_l = \sqrt{\epsilon \exp \left[ W_{-1}\left( - \frac2{\alpha(2) \epsilon}\right)\right]} \xrightarrow{\epsilon \rightarrow 0} 0
\end{equation}

\subsection{Estimation of $\omega_l (\epsilon)$ for $\mathscrsfs{D} \geq 3$}

For $\mathscrsfs{D} \geq 3$, following Eq. \eqref{kummerasymptote} we arrive at:
\begin{equation}
    \frac{ \hyperu \left( \frac{\mathscrsfs{D}-\omega_h}4+1,\frac{\mathscrsfs{D}}2+1,\omega_h\right)}{ 
\hyperu\left( \frac{\mathscrsfs{D}-\omega_h}4,\frac{\mathscrsfs{D}}2,\omega_h \right)}\Bigg|_{\mathscrsfs{D} \geq 3} \xrightarrow{\omega_h \to 0}  \frac{2(\mathscrsfs{D}-2)}{\mathscrsfs{D}} \omega_h^{-1} \ .
\end{equation}
This and Eq. \eqref{RHS_D=3} simplify Eq. \eqref{matching_for_omega_l}:
\begin{equation}
     \omega_l \left[ 1 + \frac{1}{2} (\omega_l - \mathscrsfs{D}) \frac{d}{d\zeta} \laguerrel_{\zeta}^{\mathscrsfs{D}/2 } \left(\mathscrsfs{D} \right)\Big|_{\zeta = -1} \right] = \mathscrsfs{D} - 2 \ .
\end{equation}
This equation is quadratic in $\omega_l$, which can be solved to obtain:
\begin{equation}
    \omega_l = \left( \frac{\mathscrsfs{D}}{2} - \frac{1}{Y} \right) \left[1 + \sqrt{1 + \frac{2 (\mathscrsfs{D} - 2) Y}{\left(1 - \frac{\mathscrsfs{D}}{2} Y\right)^2}} \right] \ , \ Y = \frac{d}{d\zeta} \laguerrel \left(\zeta,\mathscrsfs{D}/2, \mathscrsfs{D} \right)\Big|_{\zeta = -1} \ .
\end{equation}
This gives the asymptotic limit $\omega_l \to \mathscrsfs{D}$ as the number of traits becomes extremely large $\mathscrsfs{D} \to \infty$.

These analysis has been investigated extensively on Mathematica 13.2.

\section{The Simulation of Population Distribution on the Landscape 
\label{detail_simulation}}

We employ a random-walk agent-based simulation of stress-induced mutagenesis evolution on a $\mathscrsfs{D}=1$ dimensional landscape, in which $x \in [-L,+L]$ and the large-scale cut-off is at $L=20$. We investigate the maximum growth rate $R_0=1$, the fit region of size $\lambda=1$, the lower diffusivity $D_l=1/6$. The time resolution of our simulation is $dt=0.01$ and we ran a total time $T=100$, where the stationary state has always been reached by the end of the first half of the run. The simulation data are obtained by averaging the recorded every time interval $\Delta t=1$ at the later half of the run.

The number of agents are dynamical, an agent can be added or removed randomly with some probability $|p(x,t)|$ per dt, adding when $p(x,t)$ is positive, removing when $p(x,t)$ is negative. This $p(x,t)$ value is associated with the growth rate $G(x,t)$ at its position $x$ on the landscape. The carrying capacity is $K=10^5$ and we start with $B(t=0)=10^4$ agents randomly distributed:
\begin{equation}
    p(x,t) = G(x,t) dt \ , \ G(x,t) = R_0 \left[ 1 - \left(\frac{x}{\lambda} \right)^2 \right] \left[ 1 - \frac{B(t)}{K}\right] \ .
\end{equation}
The walking step per $dt$ of an agent is selected randomly from a Gaussian distribution centered at $0$ with a standard deviation $W(x,t)$ associated with the diffusivity $D[G(x,t)]$ at its position $x$ on the landscape:
\begin{equation}
    W(x,t) = \sqrt{2D[G(x,t)]dt} \ , \ D[G] = D_l \left[ 1 + \epsilon \Theta (-G) \right] \ .
\end{equation}
Here we investigate the stress-induced strength $\epsilon = 0, 3, 10$.

The code for this simulation is available from the corresponding author upon request. We have executed it on MatLab 2020b.






\end{document}